\documentstyle[aps,epsf,prl]{revtex}

\begin{document}
\title{
Observation of coherent $\pi^0$
electroproduction on deuterons at large momentum transfer }


\def\sphn{$^2$}
\def\cern{$^6$}
\def\jlab{$^5$}
\def\lns{$^{1}$}
\def\ncat{$^{4}$}
\def\ipn{$^{3}$}
\def\yerevan{$^{7}$}
\def\metu{$^{8}$}
\def\kfti{$^{9}$}
\author{E.~Tomasi-Gustafsson,\lns$^,$\sphn\    
	L.~Bimbot,\ipn\     
	S.~Danagoulian,\ncat$^,$\jlab\   
	K.~Gustafsson,\cern\         
	D.~Mack,\jlab\    
	H.~Mrktchyan,\yerevan\ 
	and M. P.~Rekalo\lns$^,$\metu$^,$\kfti \\
       }
\address{
\lns LNS-Saclay,  91191 Gif-sur-Yvette, France\\
\sphn DAPNIA/SPhN, CEA/Saclay, 91191 Gif-sur-Yvette, France\\ 
\ipn IPNO, IN2P3, BP 1, 91406 Orsay, France\\
\ncat North Carolina A. \& T. State University, Greensboro, NC 27411, USA\\
\cern CERN/EP, CH-1211 Geneva 23, Switzerland\\
\jlab Thomas Jefferson National Accelerator Facility, Newport News, VA 23606, 
USA\\
\yerevan Yerevan Physics Institute, 375036 Yerevan, Armenia \\
\metu Middle East Technical University,
Physics Department, Ankara 06531, Turkey\\
\kfti National Science Center: KFTI,
 310108 Kharkov, Ukraine.
}

\maketitle
\begin{abstract}
The first experimental results for coherent $\pi^0$-electroproduction on the deuteron, $e+d\rightarrow e+d +\pi^0$, 
at large momentum transfer, are reported. The experiment was performed at Jefferson Laboratory at an incident electron energy of 4.05 GeV. A large pion production yield has been observed in a kinematical region for 1.1$<Q^2<$1.8 GeV$^2$, from threshold to 
200 MeV excitation energy in the $d\pi^0$ system. The $Q^2$-dependence is compared with theoretical predictions.
\end{abstract}


\narrowtext

\section{Introduction}
After elastic scattering, the reaction 
$e+d\rightarrow e+d +\pi^0$ is the 
simplest coherent process for $ed$-collisions, which contains information on deuteron structure and on the elementary  nucleon amplitudes. This reaction has characteristics which make it a very good 
source of knowledge on the deuteron structure, complementary to other probes.
The presence of 
a deuteron with zero isospin in the initial and final states leads to a 
specific isotopic structure for the corresponding amplitudes.  
Elastic $ed$-scattering is essentially determined by an isoscalar 
combination of nucleon electromagnetic form factors, whereas coherent pion 
electroproduction on the deuteron allows a scan of the full 
isospin structure of 
the nucleon electromagnetic current in the resonance region and a 
separation of isoscalar and isovector contributions. This increases the degree of complexity, but simultaneously opens a unique way to progress in the understanding of the deuteron and nucleon structure. 

Besides the calculation of static hadronic properties like 
masses or magnetic moments, the description of elastic and inelastic form factors (which contain dynamical information on hadron structure) represents a powerful test for theoretical models. In particular, in the 
intermediate energy region, the electromagnetic form factors should be a helpful signature of the transition from the confinement regime to perturbative 
QCD \cite{Br73}. 

The experimental determination of the three form factors of the deuteron requires the measurement of polarization observables. The most recent elastic scattering $ed$ data at large values of momentum transfer squared, $Q^2\le 6$ GeV$^2$ \cite{Al99}, have been successfully compared to pQCD predictions, whereas polarization measurements \cite{Ab00} lead to the conclusion that, up to $Q^2\simeq$ 2 GeV$^2$, the deuteron structure can be described by conventional models based on nucleon and meson degrees of freedom. It seems  very hard to extend such measurements to higher momentum transfer with existing techniques \cite{Gilman}. However, inelastic processes, 
such as $e+d\to e+d+\pi^0$ (accessible with existing beams and polarimeters), probe the deuteron at smaller distances than in elastic scattering, for the same value of $Q^2$. An appropriate choice of kinematics can lead to new and interesting information. Two $d\pi^0$-excitation energy regions seem particularly promising: the threshold region 
and the $\Delta$-excitation region.

A considerable experimental activity has been going on in the 
field of near threshold pion production in $\gamma p$ collisions \cite{Fu96,Be96}, $ep$ collisions \cite{Br97}, and $\gamma d$ collisions \cite{Ber98}, but data on pion production in $ed$ collisions are scarce.  The experimental 
study of this reaction is now possible, at Mainz and at Jefferson Laboratory (JLab), due to the high duty cycle of the electron machines. Threshold $\pi^0$-electroproduction on protons and deuterons has been investigated by the A1-collaboration at Mainz \cite{Be96_2} at small momentum 
transfer squared $Q^2\le 0.1$ GeV$^2$. 

At this low momentum transfer different nonperturbative QCD approaches, in particular Chiral Perturbation Theory (ChPT), can be applied. Phenomenological approaches such as effective Lagrangians models, isobar models, quark models, and hybrid models are widely applicable. A general theoretical study of pion 
electroproduction on  deuterons was firstly developped in Ref. \cite{Re97}. An adequate dynamical approach to pion electroproduction  has to take into account 
all previous theoretical findings related to other electromagnetic processes on the deuteron, such as elastic $ed$ scattering, $\pi^0$-photoproduction $\gamma+d\to d+\pi^0$ \cite{Me99}, and  deuteron photodisintegration $\gamma+d\to n+p$ \cite{Bo98}. Like these 
processes, the reaction $e+d\to e+d+\pi^0$ involves the 
study of the deuteron structure and of the reaction mechanism, and the determination of the neutron elementary amplitudes, $\gamma^*+n\to n+\pi^0$, where $\gamma^*$ is a virtual photon. Note that the exact cancellation of rescattering effects, due to the processes: $\gamma^*+d\to p+p+\pi^-(n+n+\pi^+)\to d \pi^0$ in the near threshold region \cite{Re02}, allows to extract the neutron amplitudes from the data about $e+d\to e+d+\pi^0$, in the frame of impulse approximation.

Here the first experimental observation of $\pi^0$-electroproduction on deuterons at large values  of the momentum 
transfer squared and at relatively small excitation energy of the produced $d\pi^0$-system is reported. This kinematical region is only accessible at JLab.

The data were collected during the $t_{20}$-experiment, the primary aim of which was the measurement of the deuteron tensor polarization in elastic $ed-$scattering \cite{Ab00}. We have  reconstructed the dependence on the kinematical variables which contain the physical information for the process $e+d\to e+d+\pi^0$, taking into account the  difficulties related to limited experimental acceptance and to low detection efficiency. With a complete measurement of the five-fold differential cross section, the comparison to the theory would have been straightforward. Here the detection 
of the deuteron, due to limited resolution and statistics, does not allow a 
complete and precise reconstruction of the physical event. We derive from the experiment the $Q^2$-dependence of the yield, which can be directly compared to theoretical models, such as an effective Lagrangian model and pQCD predictions.

Our paper is organized as follows. In Section II we present kinematical and dynamical characteristics of the process $e+d\to e +d+\pi^0$. The description of the experiment is presented in Section III. Section IV is devoted to the discussion of the experimental results and to the comparison with theoretical predictions. The main results are summarized in the Conclusion. The Appendix contains a detailed scheme of the experimental analysis, in case of uncomplete event reconstruction. 

\section{The process $\lowercase{e}+\lowercase{d}\to \lowercase{e}+\lowercase{d}+\pi^0$}

\subsection{The kinematics}

In the framework of the one 
photon mechanism, the  process $e+d \to e+X$ is 
equivalent to  $\gamma^*+d \to X$, where $X$ is a hadronic system, and this gives the most convenient choice of kinematical variables for the electroproduction processes. The detection of the 
recoil deuteron in coincidence with the scattered electron allows a full reconstruction of the kinematics for $\gamma^*+d\rightarrow d+\pi^0$. 

In the limit of zero electron mass, the momentum transfer squared from the incident to the outgoing electron, $Q^2$, is defined as 
$$ Q^2=4E_1E_2 \displaystyle \sin^2\frac{\theta_e}{2},$$ 
where $E_1$, $(E_2)$ is the energy of the incident (scattered) electron and  $\theta_e$ is the electron scattering angle (in the LAB-system). As defined here, $Q^2$ is positive in the space-like region.

The energy and the angle of the scattered electron enable the determination of the momentum transfer squared, $Q^2$, and of the invariant mass of the produced hadronic system, $W$:
$$ W = \sqrt{M^2-Q^2+ 2\nu^*},$$ 
where $M$ is the deuteron mass and the quantity $\nu^*=M(E_1-E_2)$ is related to the energy transferred from the electron to the hadronic system $X$.

Events from elastic scattering and electroproduction of 
one and two pions follow straight lines in a plane defined by $Q^2$ versus $\nu^*$ (Fig. 1), corresponding to a definite value of the invariant mass $W$. Fixed values of $\theta_e$ correspond also to straight lines in the $Q^2$-$\nu^*$ plane. In Fig. 1 the lines corresponding to 
$\theta_e=18.5^{\circ}(\pm 1.5^{\circ})$, are drawn to emphasize the kinematical limits
of the experimental set-up.

For the analysis of the  $\pi^0$ production data near threshold, instead of the invariant variable $t=(p_1-p_2)^2$ ($p_1$ and $p_2$ are the four-momenta of the target and of the outgoing deuteron, respectively), it is preferable to use  $\cos \tilde{\theta_\pi}$, where $\tilde{\theta_\pi}$ is the pion production angle 
in the center of mass system (CMS) of the final $d\pi^0$-system . 

At a given value of $W$, the final deuteron energy in the laboratory system, $E_d$, can be expressed as a quadratic 
function of the cosine of the deuteron scattering angle, $\cos \theta_d$, which is drawn in Fig. 2 for different values of $W$. In this figure the threshold point for $\pi^0$ electroproduction and the point for elastic $ed$ kinematics, at a fixed value of incident electron energy and electron scattering angle, are also indicated.

These considerations are valid for coplanar kinematics, where all momenta of the
final particles in $e+d\rightarrow e+d+\pi^0$ are in the same plane. This
corresponds to two values of  the azimuthal angle of the final deuteron,
$\phi=0$ and $\phi=\pi$, relative to the electron
scattering plane, defined by the directions of the three-momenta of the initial and
final electrons $\vec {k_1}$ and $\vec {k_2}$. The left-hand side of each
ellipse in Fig. 2 (with respect to the center, i.e. the threshold point)
corresponds to $\phi=\pi$ (the deuteron scattering angle is smaller than the
threshold value) and the right-hand side of the ellipses corresponds to
$\phi=0$ (the deuteron scattering angle is larger than the threshold value).

Note, in conclusion, that the  measurement of $E_d$ allows 
the determination of  $\cos \tilde{\theta_\pi}$ as a function of $E_d$, $W$, and $Q^2$ through the following expression:
\begin{eqnarray}
&2M^2-2ME_d=&-Q^2+m_{\pi}^2-
\displaystyle\frac{(W^2-Q^2-M^2)(W^2+m_{\pi}^2-M^2)}{2W^2}
\\&& +2cos \tilde{\theta_\pi}\nonumber
\sqrt{\left [\displaystyle\frac{(W^2-Q^2-M^2)^2}{4W^2}+Q^2\right ]\left [\displaystyle\frac {(W^2+m_{\pi}^2-M^2)^2}{4W^2}-m_\pi^2 \right ]}.
\end{eqnarray}
The knowledge of the $\cos \tilde{\theta_\pi}$-dependence for the differential cross section of $\gamma^*+d\to d+\pi^0$, is essential in order to perform a multipole analysis.

\subsection{The dynamics}

In the framework of the one-photon mechanism, the differential cross section for $e+d\rightarrow e+d +\pi^0$, can be written as \cite{Ak77}:
\begin{eqnarray}
\sigma(\phi)=\frac{d^5\sigma}{dE_2d\Omega 
\widetilde{d\Omega}}={\cal N}\left[{\cal H}_{T}+
\epsilon {\cal H}_{L}+\epsilon{\cal H}_{P}\cos 
2\phi+\sqrt
{2\epsilon(1+\epsilon)}
{\cal H}_{I}\cos\phi\right ],
\label{eq:par1}
\end{eqnarray}
where ${\cal N}$ is a normalization kinematical coefficient: 
\begin{equation}
{\cal N}=\displaystyle\frac{\alpha^2}{64\pi^3}\frac{E_2}{E_1}
\frac{q_\pi}{MW}\frac{1}{(1-\epsilon)}\frac{1}{Q^2},
\label{eq:nor}
\end{equation}
and $\epsilon$ is the degree of linear polarization of the virtual photon:
\begin{equation}
\epsilon^{-1}=1+2\displaystyle\frac{\vec 
k_{\gamma}^2}{Q^2}\tan^2\displaystyle\frac{\theta_e}{2},
\end{equation}
Here $\vec k^2_{\gamma}=(\vec k_1-\vec 
k_2)^2=E_1^2+E_2^2-2E_1E_2\cos\theta_e$, $\vec{q_\pi}$ is the pion three-momentum in the CMS of the reaction $\gamma^*+d\rightarrow d+\pi^0$ with $\vec q^2_{\pi}=E^2_{\pi}-m^2_{\pi}$, $E_{\pi}= \displaystyle\frac{W^2+m_{\pi}^2-M^2}{2W}$, and $\alpha=\displaystyle\frac{e^2}{4\pi}\simeq\displaystyle\frac{1}{137}$.

The terms ${\cal H}_{a}$ ($~a=T,L,P, I$), are related to the four standard 
contributions to the differential cross section for $\gamma^*+d\rightarrow d+\pi^0$, corresponding 
to the different polarizations of the virtual photon: $T$, $P$ are the 
transverse components, $L$ is the longitudinal component and $I$ is the interference between the longitudinal and the transversal components. The element of solid angle for the scattered electron (deuteron) in the LAB (CMS) system is ${d\Omega}$ ($\widetilde{d\Omega}$). Note that $\widetilde{d\Omega}=d\cos{\tilde\theta_\pi} d\phi_\pi$.

The different contributions ${\cal H}_{a}$ depend 
on $Q^2,~W,$ and $\cos\tilde{\theta_\pi}$. The azimuthal dependence is explicit in the cosine terms, Eq. (2). The three kinematical quantities $\epsilon$, ${\cal N}$ and $\phi$ depend on the electron kinematics. 

A measurement of $\sigma(\phi)$ for 3 values of $\phi$ (for example $\phi=0$, $~\pi/2$,  and $~\pi$)  and for two values of the parameter 
$\epsilon$ allows a complete and unique separation of all the four contributions to the cross section. Note that Eq. (\ref{eq:par1}) can be considered a generalization of the Rosenbluth formula for a three-body reaction, in which only two particles are detected in the final state.

In the near threshold region, in the framework of the S- and 
P-waves pion production, the four contributions to the differential cross section of 
$e+d\to e+d+\pi^0$, can be parameterized as the functions of 
$\cos \tilde{\theta_\pi}$ as follows (omitting, for simplicity, the deuteron form factors):
\begin{eqnarray}
&{\cal H}_{T}=a_0+a_1\cos 
\tilde{\theta_\pi}+a_2\cos^2 \tilde{\theta_\pi},~~~{\cal H}_{P}=b_0\sin^2 
\tilde{\theta_\pi},\nonumber\\
&{\cal H}_L=c_0+c_1\cos 
\tilde{\theta_\pi}+c_2\cos^2 \tilde{\theta_\pi},~~~
{\cal H}_I=\sin \tilde{\theta_\pi}(d_0+d_1\cos 
\tilde{\theta_\pi}),\label{eq:par2}
\end{eqnarray}
where the real coefficients $a_i$, $b_i$, $c_i$, and $d_i$ are well defined quadratic combinations of multipole amplitudes for $\gamma^*+d\to d+\pi^0$, which are functions of only two variables, $Q^2$ and $W$. All the dynamical information about this process is contained in these multipole amplitudes. The  experimental determination of the $Q^2$ and $W$ dependences of the multipole amplitudes would allow a direct comparison with the theory. In the 
framework of the S- and P-wave contributions, the five-fold cross section 
has to be measured for at least nine points (for different $\cos\tilde{\theta_\pi}$, $\phi$, and  $\epsilon$) in order to fully determine the multipole amplitudes (the moduli and relative phases).

In the case of limited acceptance or of partial information on one or both of the final particles, one can extract from the experiment - and compare to  
theoretical predictions - only some combinations of the above mentioned coefficients. For example, near threshold, the pions are emitted in a narrow cone around the virtual photon direction (in the LAB system) and the experimental resolution may not allow a precise determination of the azimuthal angle, or, on the contrary, far above threshold, the detection acceptance may not cover the full phase space. We detail in the Appendix, a rigorous method for the data analysis, in case of limited kinematical information.

\section{The experiment}

From Figs. 1 and 2, it appears that the kinematical characteristics of the 
outgoing particles (the scattered electron and deuteron) in the  process 
$e+d\rightarrow e+d+\pi^0$ in the threshold region, are near to those of the 
elastic process $e+d\rightarrow e+d$. Therefore the experimental set up of the $t_{20}$ experiment at JLab, which had a double arm detection 
for elastic $ed$ scattering,  could be used to  study the inelastic process 
of $\pi^0$-production. The experiment was performed in Hall C. With small 
changes in the  spectrometer settings, it was possible to reach near threshold  kinematics in which $\pi^0$ events were detected. The typical luminosity  was about 2$\cdot$10$^{38}$ cm$^2$s$^{-1}$ 
obtained with a 40 $\mu$A continuous electron beam and a 12 cm long liquid 
deuterium target. The electrons were detected in a  
large solid angle (6 msr) spectrometer (HMS) with an energy resolution $\Delta 
E/E=10^{-3}$. The deuterons were focussed on the polarimeter POLDER \cite{Ko94} 
through a magnetic transport line located at a fixed angle, $\theta_d$, of  $60.5^\circ$. For 
the initial electron energy, $E_e=$4.05 GeV, 
the scattered electrons were detected at an angle $\theta_e=18.5^\circ$
corresponding  to a range of four momentum squared 1.1$<Q^2<$1.8 
GeV$^2$.
The coincidence between 
electrons and deuterons reduced the high background. A more 
detailed 
description of the experimental set up can be found in \cite{Ab00,Ab99}.

The deuterons were identified  in the two-dimensional spectrum 
corresponding to time of flight versus the ADC signal related 
to the energy loss in the POLDER start detectors. 
The deuterons were selected  by the contour shown in Fig. 3 (contour I, labelled 'signal'). An estimation of the background, done by 
displacing the contour (contour II, labelled 'background'), is about 20\%. The largest part of the background, corresponding to protons coming from deuteron break up does not appear on the figure, as it corresponds to a different region of the time of flight spectrum. 

The  spectrum of the invariant mass W is shown in Fig. 4 for different selection criteria of events.
Above the pion threshold, $W_{th}=M+m_\pi$, a significant number of events were observed. The transition between the 
elastic and the pion production regions is illustrated in Figs. 5 and 6. The 
number of counts is plotted as a two-dimensional function  of  $Q^2$  and $W$, for the 
$e+d\rightarrow e+ X$ reaction, for spectrometer settings corresponding to  
elastic 
kinematics (Fig. 5) and to pion kinematics (Fig. 6), where a tail from elastic 
scattering is still visible. The experimental resolutions are
$\displaystyle\frac{\Delta W}{W}=0.3\% \mbox{ and }\displaystyle\frac{\Delta 
Q^2}{Q^2}=0.7\%$.

In this measurement, in four hours beam on target, 25815  pion events were counted in contour I, 
330 of which correspond to the near 
threshold region in an invariant mass range 
$\Delta W=W-W_{th}$=40 MeV. 
This total number  is comparable to 
the number of events 
for elastic $ed$-scattering in similar experimental conditions. 

\section{The results}

The deuteron magnetic channel has large angular acceptance and low momentum resolution. The deuteron momentum and scattering angle could not be reconstructed with precision. For this reason the information presented here concerns the kinematical variables calculated from the electron channel. We focus here on the $Q^2$-dependence of the differential cross section of the process $e+d\to e+d+\pi^0$, for which theoretical predictions are available.

In Fig. 7 we show the Monte Carlo expectation \cite{Pitz} for the $Q^2$ and the $W$-distributions, calculated for a uniform input distribution (solid line) and for a distribution weighted by the kinematical factor ${\cal N}$ (Eq. \ref{eq:nor}). The figure shows the range of detection, where the acceptance of the apparatus is reasonably flat:
1.3 $\le Q^2\le$ 1.6 GeV$^2$ and  $W$ from threshold up to 2.3 GeV.

In order to extract the $Q^2$-distribution, quite conservative 
cuts were applied in order to select events well inside the acceptance of the deuteron channel. We assumed that the efficiency is constant in this central region. This is reasonably supported by the Monte Carlo simulations.

Systematic errors due to event selection were estimated with the help of a parallel analysis, where the selection of events was done by a window in the time of flight and electron momentum bi-dimensional plot. Background subtraction was done by displacing a window in the time of flight spectrum. The final distributions were consistent within the error bars.

In Fig. 8 the
$Q^2$-dependence of the counting rates integrated 
over the experimental acceptance, is given for 
different region of W, in bins of 40 MeV, from threshold to the $\Delta$-excitation region. The data are corrected by the kinematical factor $\cal N$ (see Eq. \ref{eq:nor}) in order to make an easier comparison with the theoretical predictions. We did not attempt to apply an absolute normalization of the data due to a too large uncertainty on the  reconstruction of the deuteron kinematics, although the $d\pi^0$ events were unambiguosly identified. 

Moreover the radiative corrections are necessary to extract absolute values of the cross section.  For the considered process, at relatively large momentum transfer, radiative corrections are, in principle, far from being negligible. Their calculation is complicated from a theoretical point of view, and at our knowledge, no calculation exists for pion coherent electroproduction on the deuteron. Furthermore, the  acceptance has to be taken precisely into account. But, if we consider relative yields, this problem can be neglected, mainly due to the logarithmic, i.e. weak dependence of radiative correctionson $Q^2$ and relatively small $Q^2$-interval in the present  experiment. 

The four spectra present a similar steep decreasing behavior \footnote{The deviation from a monotonic decreasing for $Q^2\ge 1.6$ GeV$^2$ may reflect a limitation in the acceptance (see Fig. 7).}. In general the acceptance in one variable is a complicated function of other variables and it is usually estimated through sophisticated simulations. But if the $\phi$-acceptance is small or constant, or in the case of full 
$2\pi$ $\phi$-acceptance, a rigorous treatment of the data is possible, even 
without full information on the azimuthal angle (see Appendix). 
The relative $Q^2$-dependence of the yields can be compared with theoretical 
calculations, such as impulse approximation or pQCD scaling laws \cite{Ma73}.

Predictions available from a classical (mesonic) model on coherent pion electroproducion \cite{Re97}, where the reaction mechanism is described within the impulse approximation (the deuteron is described by the Paris wave function), and the $\gamma^*+N\to N+\pi$-interaction is treated in the framework of an effective Lagrangian model \cite{Re97}, are shown in Fig. 8.  The solid and dashed lines correspond respectively to $\phi$-integration over $2\pi$ and to the limit for small $\Delta\phi$, which is closer to the experimental conditions. For $2\pi$-acceptance only ${\cal H}_{T}$ and $ {\cal H}_{L}$ contribute to the cross section, whereas, in the case of small $\phi$-acceptance, all contributions to the exclusive cross section are present (see Appendix).
 
The theoretical curves are normalized to the highest experimental point, for the smallest value of $Q^2$. After normalization, the results are not very sensitive to the opening of the azimuthal angle. Such behavior can be interpreted as an indication of a weak $\phi$-dependence of the $d(e,e'\pi^0)d$ cross section. Another possibility is that the different contributions induce a similar $Q^2$-dependence of the cross section, integrated in this kinematical region. The theoretical model \cite{Re97} predicts indeed a large $\phi$-dependence in the $\Delta$-region.

Following the quark counting rule of pQCD, \cite{Br73}, the asymptotic behaviour of the 
electromagnetic  (elastic and inelastic) form factors of hadrons follows a 
$(1/Q^2)^{(n_1+n_2)/2-1}$ dependence, where $n_1(n_2)$ is the number of quarks 
in the initial (final) state.
For pion electroproduction on the deuteron (at relatively large momentum transfer and small excitation energies, where the electroproduction process is determined by the inelastic electromagnetic current, $\gamma^*+d\to d+\pi^0$,  with $n_1=6$ and $n_2=6+2=8$) we expect a value of $n_1+n_2=14$, 
which corresponds to a steeper decrease of the cross section, compared to elastic $ed-$scattering, where $n_1+n_2=12$.  

The $Q^2$-behavior for coherent 
inelastic deuteron cross section is illustrated in Fig. 8,
where we show the results of the parametrization:
\begin{equation}
\sigma_{d\pi^0}(Q^2)=\displaystyle\frac{\sigma_{d\pi^0}(0)}{\left( 
1+\displaystyle\frac{Q^2}{m^2}\right 
)^N},
\end{equation}
for  $N=14$ (dotted line), and for  $m^2$=1.41 GeV$^2$, according to \cite{Br73}.

The results from these two approaches are consistent with the present data. As for elastic $ed$-scattering \cite{Al99}, the measurement of the cross section alone does not allow us to disentangle predictions given from different models of the deuteron structure.

\section{Conclusions}

Coherent $\pi^0$ electroproduction on the deuteron, 
$e+d\rightarrow e+d +\pi^0$, at 
relatively large momentum transfer,
has been detected for the first time. 
The specific conditions of this experiment covered coherent $\pi^0$-production 
in the near threshold region and in the region of excitation of the $\Delta $-resonance. 
A steep decrease with $Q^2$ of the counting rate has been observed at different 
values of $W$.

The present results show that it is possible to foresee a research programme based on the experimental study of coherent pion electroproduction on the deuteron at relatively large momentum transfer, at threshold and in the $\Delta$-region, to access:
\begin{itemize}
\item the relative contributions of $S$- and $P$-waves for different values of 
$Q^2$,
\item the  $Q^2$-scaling behavior of $S$- and $P$-waves excitation 
for $\gamma^*+d\rightarrow d+\pi^0$,
\item the specific mass parameter, $m^2$, which enters in the $Q^2$ dependence of the different contributions to the differential cross section to be compared to  meson and nucleon form factors values, and 
\item the $\Delta$-isobar excitation on the deuteron, $\gamma^*+d\to \Delta+N\to d+\pi^0$. 

\end{itemize}

We have established the feasibility of such an experimental study, since counting rates are similar to elastic scattering.
More complete results could be obtained 
at the Jefferson Laboratory in a dedicated experiment, which would stimulate 
parallel efforts  from the theoretical side in developing specific calculations adapted to this newly accessible region. In particular, in addition to differential cross section,
measurements of the vector and tensor polarization of the outgoing deuteron are possible in this energy domain \cite{etg}.

Finally, we would like to recall that after several decades of experimental and theoretical studies of $ed$ elastic scattering, the situation with the deuteron models 
(choices of nucleon form factors, deuteron wave functions, corrections to impulse approximation...) is not yet disentangled. In this respect a detailed 
study of $e+d\to e+d +\pi^0$ will bring new important pieces of information.

\section{Acknowledgments}

We would like to thank all the members of the $t_{20}$ collaboration for help in taking the data, and, in particular, J. Arvieux, E. Beise, R. Gilman, C. Glashausser and S. Kox, for useful discussions.

\section{Appendix}

We present here a possible scheme for the analysis of 
$e^-+d\to e^-+d+\pi^0$ data, taking into account the case of partial information of the detected particles.

For the estimation of the $\phi-$acceptance it is necessary 
to 
know the relative angle between the 3-momentum of the virtual photon, 
$\vec k_{\gamma}$, and the momentum of the scattered deuteron. This angle 
depends 
on the variable W and can be calculated using the following expression for the 
production angle of the virtual photon, $\theta_{\gamma}$, relative to the 
electron scattering angle:
$$\cos\theta_{\gamma}=\displaystyle\frac{E_2\cos\theta_e+E_1}
{k_{\gamma}}.$$

\vspace*{.3 true cm}
{\noindent\bf\underline{ Limited $\phi-$acceptance}}
\vspace*{.2 true cm}

If we approximate the $\phi-$acceptance for the emitted deuteron as:
$$
-\displaystyle\frac{\Delta\phi}{2}\le\phi\le
\displaystyle\frac{\Delta\phi}{2},$$
then, for $\Delta\phi\ll 1$, all possible contributions in Eq. 
(\ref{eq:par1}), 
namely 
$1$, $\cos 2\phi$ and $\cos\phi$, will give the same results:
\begin{eqnarray*}
&\int_{-\Delta\phi/2}^{\Delta\phi/2}~1~d\phi=\Delta\phi,\nonumber\\
&\int_{-\Delta\phi/2}^{\Delta\phi/2}\cos 2\phi d\phi=\sin\Delta 
\phi\simeq\Delta\phi,\nonumber\\
&\int_{-\Delta\phi/2}^{\Delta\phi/2}\cos \phi 
d\phi=\sin\Delta\phi\simeq\Delta\phi.
\end{eqnarray*}
So we can write the $\phi$-integrated cross section as follows:
$$\displaystyle\frac{d^4\sigma}{dE_2d\Omega d\cos\tilde\theta_\pi}=\Delta\phi 
{\cal N}\sigma(Q^2,W,\cos\tilde\theta_\pi,E_1),$$
with the following dependence for $\cos\tilde\theta_\pi$:
\begin{equation}
\sigma(Q^2,W,\cos\tilde\theta_\pi,E_1)=A_0+A_1\cos\tilde\theta_\pi+A_2
\cos^2\tilde\theta_\pi+A_3 \sin\tilde\theta_\pi+A_4\sin\tilde\theta_\pi 
\cos\tilde\theta_\pi,
\label{eq:sigma4}
\end{equation}
where the five coefficients $A_i$ are definite linear combinations of the 
coefficients (\ref{eq:par2}):
\begin{eqnarray}
&A_0&=a_0+\epsilon(c_0+b_0),\nonumber\\
&A_1&=a_1+\epsilon c_1,\nonumber\\
&A_2&=a_2+\epsilon(c_2-b_0),\label{ac}\\
&A_3&=\sqrt{2\epsilon(1+\epsilon)}d_0,\nonumber\\
&A_4&=\sqrt{2\epsilon(1+\epsilon)}d_1.\nonumber
\end{eqnarray}
The $E_1$-dependence of all these coefficients $A_i$ is contained only in the 
parameter $\epsilon$.

If we can measure the cross section 
$\displaystyle\frac{d^4\sigma}{dE_2d\Omega d\cos\tilde\theta_\pi}$ at five different values of $\tilde\theta_\pi$, we will determine all the five  coefficients $A_i(Q^2,W,\cos\tilde\theta_\pi,E_1)$, which can be compared to theoretical predictions.

\vspace*{.3 true cm}
{\noindent\bf\underline{ Full $2\pi$ acceptance}}
\vspace*{.2 true cm}

In this case only the L and T components of the cross section contribute to 
the 
integral of  (\ref{eq:par1}) in the interval $0\le\phi\le2\pi$:
$$\displaystyle\frac{d^4\sigma}{dE_2d\Omega d\cos\tilde\theta_\pi}=2\pi 
{\cal N}\sigma(Q^2,W,\cos\tilde\theta_\pi,E_1),$$
with the following dependence for $\cos\tilde\theta_\pi$:
\begin{equation}
\sigma(Q^2,W,\cos\tilde\theta_\pi,E_1)=A_0+A_1\cos\tilde\theta_\pi+A_2
\cos^2\tilde\theta_\pi
\label{eq:sigma5}
\end{equation}
and the three coefficients $A_i$ are definite linear combinations of the 
coefficients (\ref{eq:par2}):
\begin{eqnarray}
&A_0&=a_0+\epsilon c_0,\nonumber\\
&A_1&=a_1+\epsilon c_1,\label{ac2}\\
&A_2&=a_2+\epsilon c_2.\nonumber
\end{eqnarray}
The Rosenbluth fit in $\epsilon$ is very useful for 
the separation of the different contributions to the coefficients $A_0-A_2$.

\subsection{$\cos\tilde\theta_\pi$-integration}

This integration has to be done, if the deuteron 
energy (in the LAB-system) is not properly measured.

The representation (\ref{eq:par2}) is well adapted to 
$\cos\tilde\theta_\pi$-integration over the energy acceptance of the deuteron 
channel. We can use the one-to-one correspondance between the deuteron energy 
$E_d$ (in LAB-system) and $\cos\tilde\theta_\pi$. From Eq. (\ref{eq:par1}) we find:
$$\cos \tilde{\theta_\pi}=\frac{T_0-E_d}{T_1},$$
where $T_0$ and $T_1$ are:
$$T_0=\frac{1}{4M}\left[
W^2-Q^2-m_\pi^2+\displaystyle\frac{(M^2-Q^2)(M^2-m_\pi^2)}{W^2}\right ],$$
$$T_1=\frac{1}{M}\left[\left(\displaystyle\frac{(W^2+Q^2-M^2)^2}{4W^2}
-Q^2\right ) \left ( 
\displaystyle\frac{(W^2+m_\pi^2-M^2)^2}{4W^2}-m_\pi^2\right )\right ]^{1/2},
$$
i.e. the energies $T_0$ and $T_1$ are functions of $Q^2$ and $W$ only.

The final result  can be written as:
$$\displaystyle\frac{d^3\sigma}{dE_2d\Omega}=\Delta\phi 
{\cal N}\tilde\sigma(Q^2,W,E_1),$$ 
where 
$$\tilde\sigma(Q^2,W,E_1)=A_0I_0+A_1I_1+A_2I_2+A_3I_3+A_4I_4.$$
The coefficients $I_0-I_4$ are the integrals over the acceptance 
of the deuteron channel:
\begin{eqnarray}
&I_0&=\int_{\Delta}d\cos 
\tilde{\theta_\pi}=\displaystyle\frac{1}{T_1}\int_{E_{d,min}}^{E_{d,max}}dE_{d
},
\nonumber\\
&I_1&=\int_{\Delta}\cos \tilde{\theta_\pi}d\cos 
\tilde{\theta_\pi}=-\displaystyle\frac{1}{T_1^2}\int_{E_{d,min}}^{E_{d,max}}
(T_0-E_d)dE_{d},
\nonumber\\
&I_2&=\int_{\Delta}\cos^2 \tilde{\theta_\pi}d\cos 
\tilde{\theta_\pi}=-\displaystyle\frac{1}{T_1^3}\int_{E_{d,min}}^{E_{d,max}}
(T_0-E_d)^2dE_{d},\\
&I_3&=\int_{\Delta}\sin \tilde{\theta_\pi}d\cos 
\tilde{\theta_\pi}=-\displaystyle\frac{1}{T_1}\int_{E_{d,min}}^{E_{d,max}}
dE_{d}\sqrt{1-\displaystyle\frac{(T_0-E_d)^2}{T_1^2}},\nonumber\\
&I_4&=\int_{\Delta}\sin \tilde{\theta_\pi}\cos \tilde{\theta_\pi}d\cos 
\tilde{\theta_\pi}=-\displaystyle\frac{1}{T_1^2}\int_{E_{d,min}}^{E_{d,max}}
dE_{d}(T_0-T_1)\sqrt{1-\displaystyle\frac{(T_0-E_d)^2}{T_1^2}},
\nonumber
\end{eqnarray}
where $E_{d,min}$ and $E_{d,max}$ are the minimal and maximal energies of the 
deuteron (in the LAB-system). The coefficients $I_0-I_4$ are functions of $Q^2$ 
and 
$W$, for each initial energy of the electron beam, $E_1$.

The calculation of the coefficients $A_0-A_4$, in the framework of a 
definite model for $\gamma^*+d\rightarrow d+\pi^0$, together with numerical 
values for $I_0-I_4$, allows a straightforward comparison of the measured 
cross section $\displaystyle\frac{d^2\tilde\sigma}{dQ^2dW}$ with theoretical 
predictions.

\begin{figure}
\begin{center}
\mbox{\epsfxsize=15.cm\leavevmode \epsffile{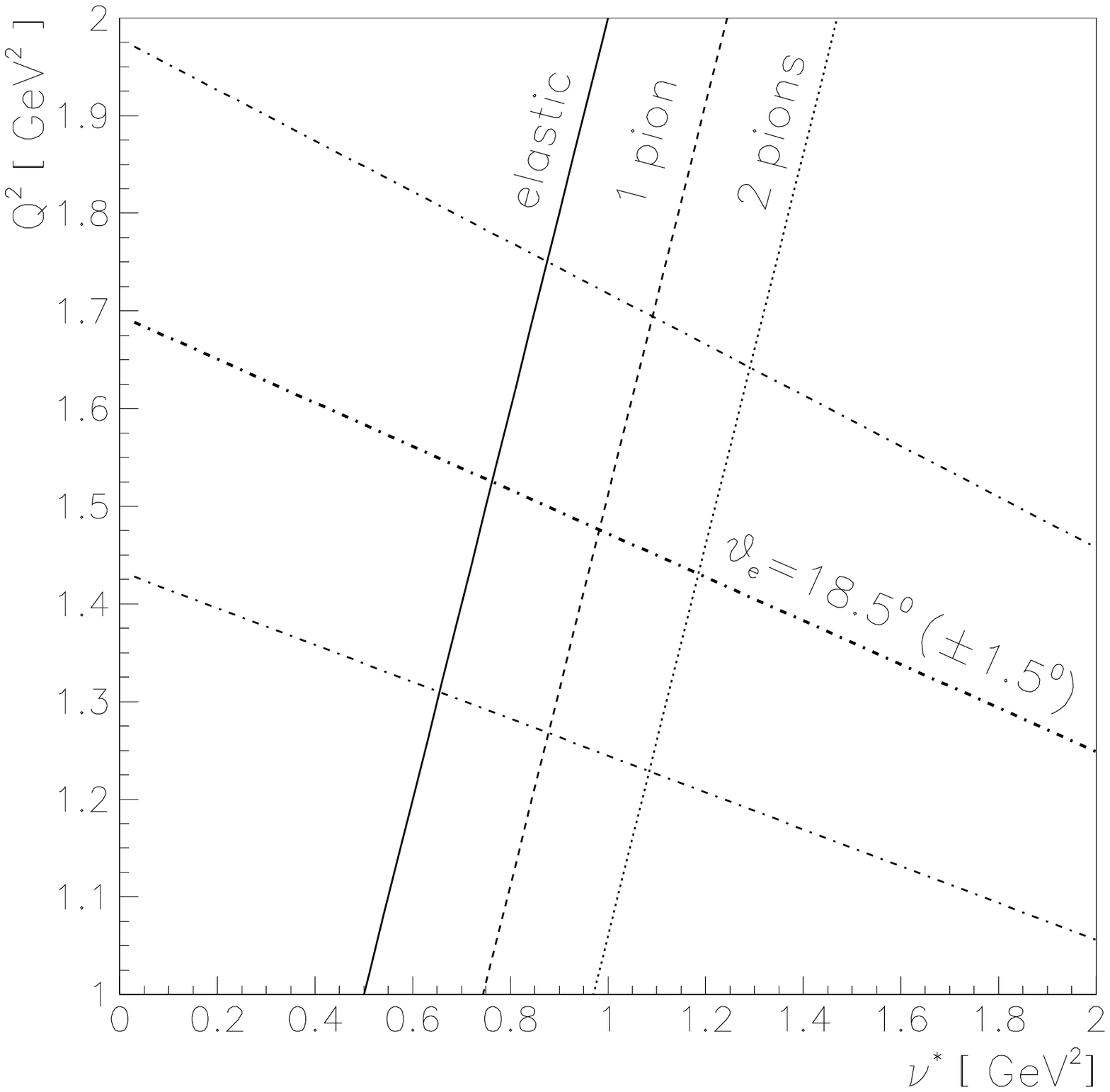}}
\end{center}
\vspace*{-2truecm}
\caption{Kinematical lines of $ed-$interaction,
in the $Q^2$ versus $\nu^*$ plane, calculated for a beam energy $E_1$=4.05 GeV. 
The elastic 
line (solid line) and the threshold lines for one pion (dashed line) and two pion production (dotted line) are 
indicated. The angular range covered by the kinematics of the electron is also 
indicated (dashed-dotted lines).}
\label{fig1}
\end{figure}
\begin{figure}
\begin{center}
\mbox{\epsfxsize=15.cm\leavevmode \epsffile{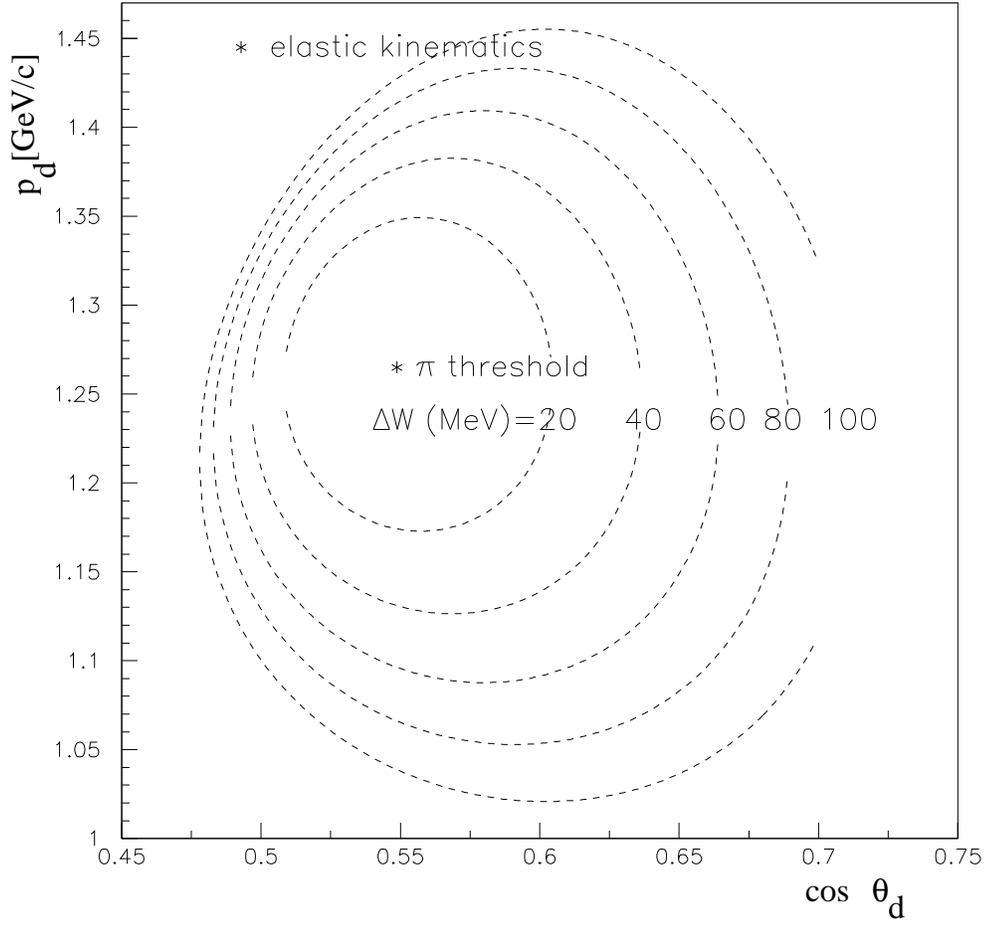}}
\end{center}
\caption{The deuteron momentum as a function of the 
deuteron scattering angle, in coplanar kinematics, for different values of 
$\Delta W$, the excess of invariant $d\pi $-mass  over the pion production threshold. 
The initial electron energy is $E_1=$4.05 GeV and $\theta_e=18.5^{\circ}$.}
\label{fig2}
\end{figure}
\newpage

\begin{figure}
\begin{center}
\mbox{\epsfxsize=15.cm\leavevmode \epsffile{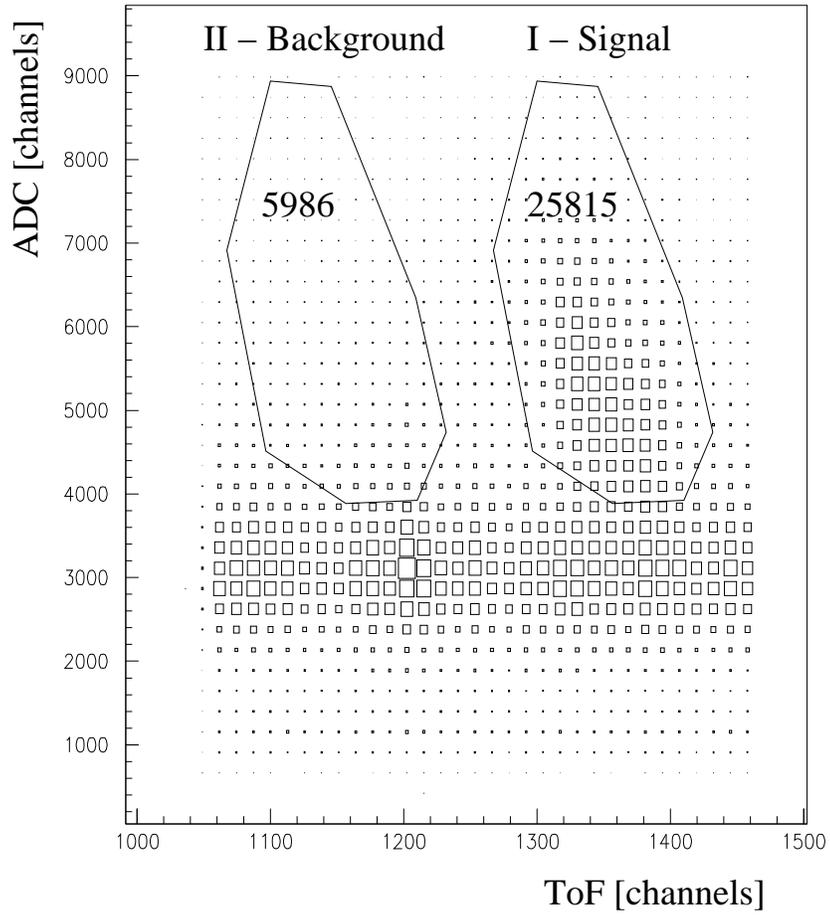}}
\end{center}
\caption{Two-dimensional plot of the events in the ADC  versus time of 
flight plane. The contours are used to select the deuterons from 
$e+d\to e+d+\pi^0$ events (contour I, labelled 'signal') and to estimate the background (contour II, labelled 'background'). The events at ADC values around 
3000 correspond to random electron-protons coincidences.}
\label{fig3}
\end{figure}
\newpage
\begin{figure}
\begin{center}
\mbox{\epsfxsize=15.cm\leavevmode \epsffile{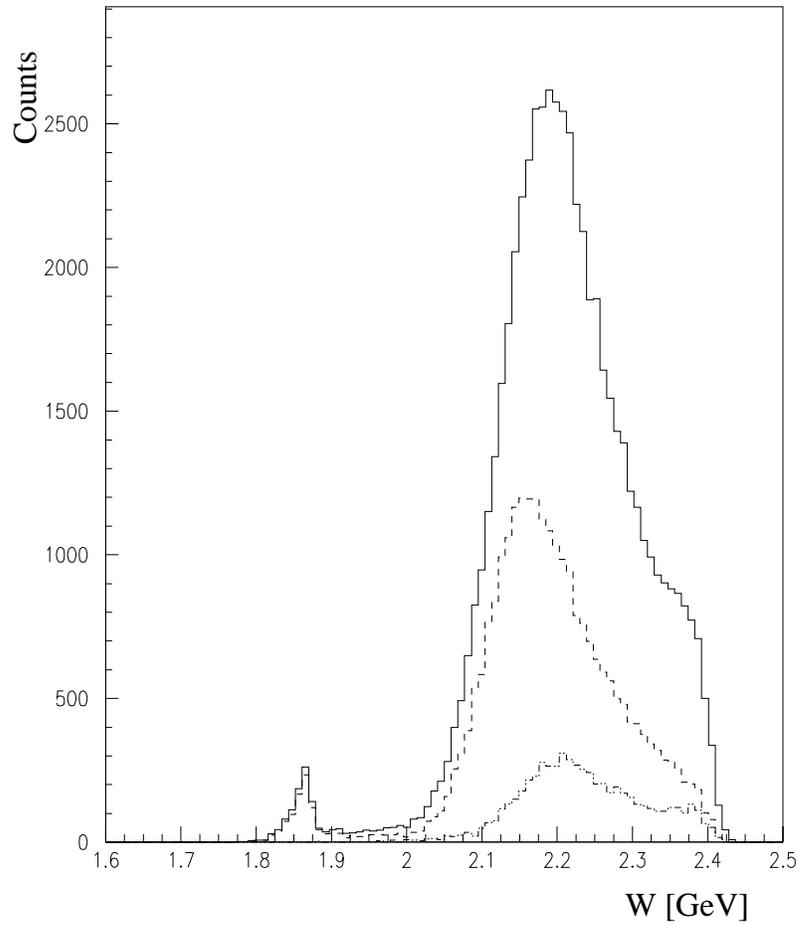}}
\end{center}
\caption{Experimental distribution of the invariant mass, $W$, corresponding 
to events in the range of time of flight 1280-1440 
(full line) and to events selected, according to Fig. 3, by  contour I (dashed 
line)  or by  contour II (dotted line).}
\label{fig4}
\end{figure}
\newpage
\begin{figure}
\begin{center}
\mbox{\epsfxsize=15.cm\leavevmode \epsffile{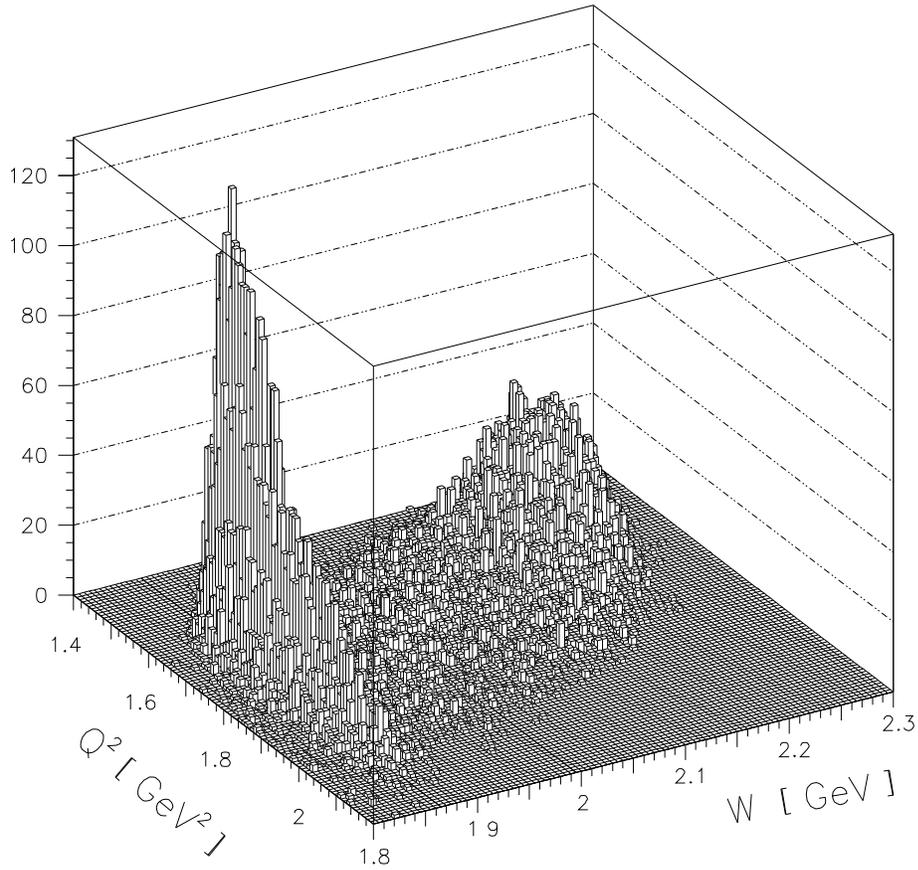}}
\end{center}
\vspace*{-3.truecm}
\caption{Two-dimensional plot of the number of counts as a function of 
momentum transfer squared, $Q^2$, and 
invariant 
mass of the hadronic  system, W, for the 
$e+d\to e+ X$ reaction,  in the near elastic 
kinematics. The beam energy is 4.05 GeV and the $\theta_e$= 20.3$^o$. Events corresponding to the elastic peak are centered around $W=M$. }
\label{fig5}
\end{figure}
\newpage
\begin{figure}
\begin{center}
\mbox{\epsfxsize=15.cm\leavevmode \epsffile{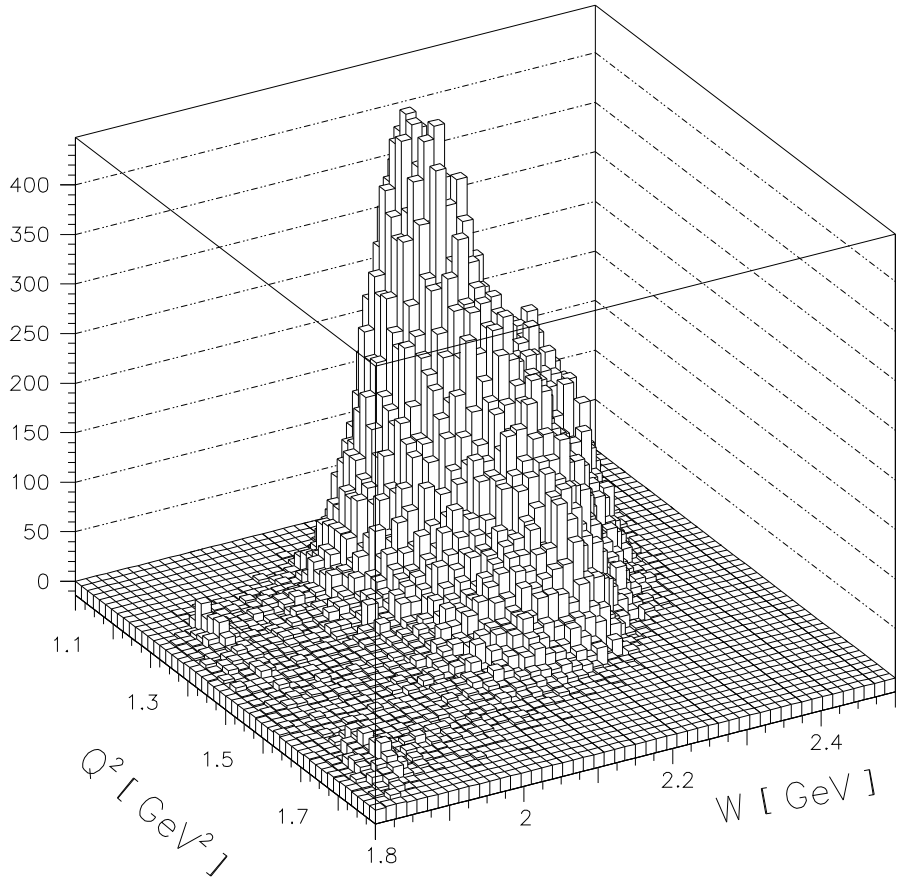}}
\end{center}
\vspace*{-3.truecm}
\caption{Two-dimensional plot of the number of counts as a function of 
momentum transfer, $Q^2$, and 
invariant 
mass of the hadronic  system, W, for the 
$e+d\rightarrow e+ X$ reaction,  in the pion 
kinematics. The beam energy is 4.05 GeV and $\theta_e$= 18.5$^o$. 
Events corresponding to pion production are 
visible for $W>M+m_{\pi}$.}
\label{fig6}
\end{figure}
\newpage
\begin{figure}
\begin{center}
\mbox{\epsfxsize=15.cm\leavevmode \epsffile{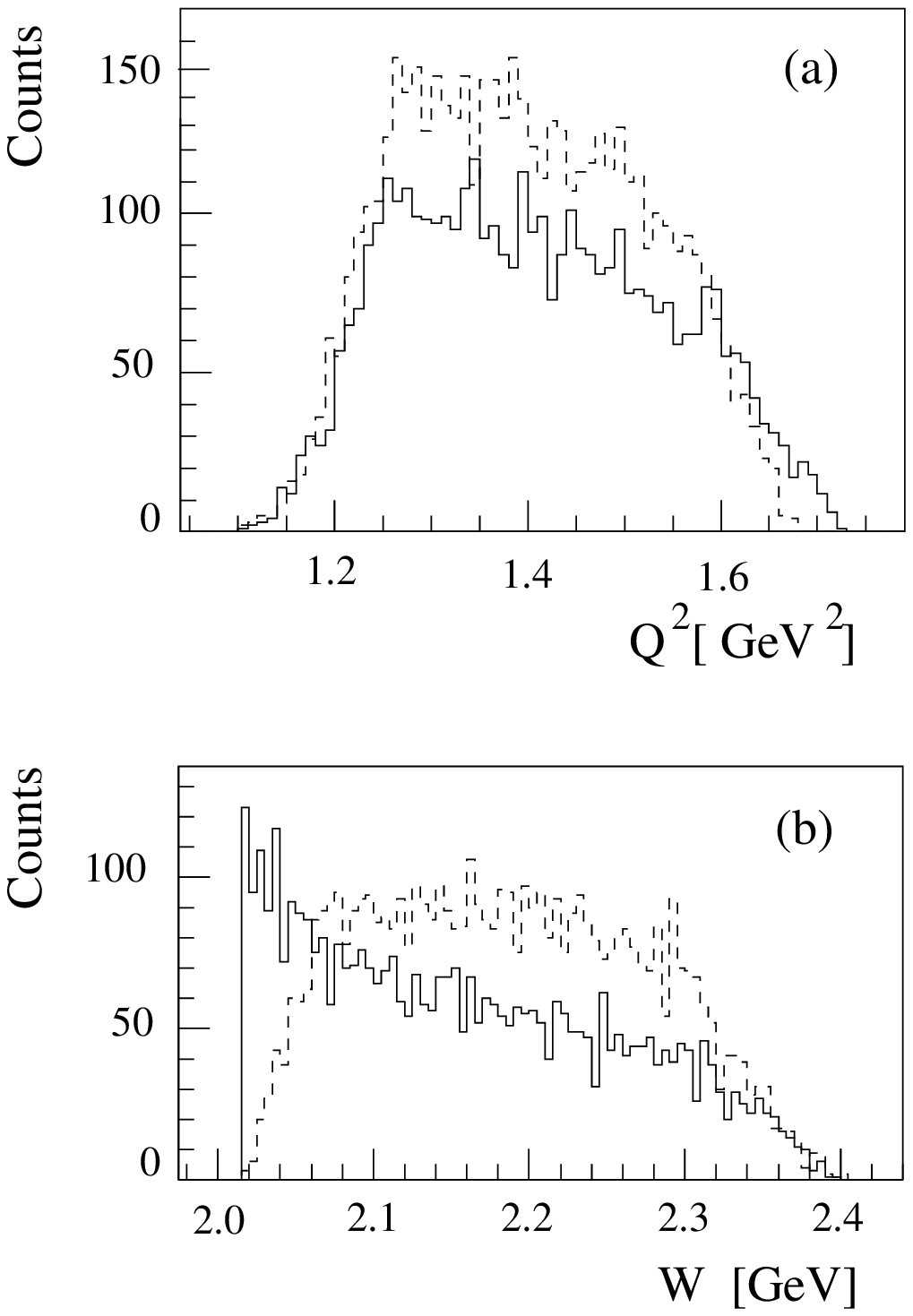}}
\end{center}
\vspace*{-3truecm}
\caption{$Q^2$-distribution (a) and $W$-distribution (b), following a Monte Carlo simulation. The full line corresponds to an uniform input cross section and
the dashed line corresponds to an input distribution corrected by the kinematical factor ${\cal N}$ (Eq. \protect\ref{eq:nor}).
}
\label{sim}
\end{figure}
\newpage
\begin{figure}
\begin{center}
\mbox{\epsfxsize=15.cm\leavevmode \epsffile{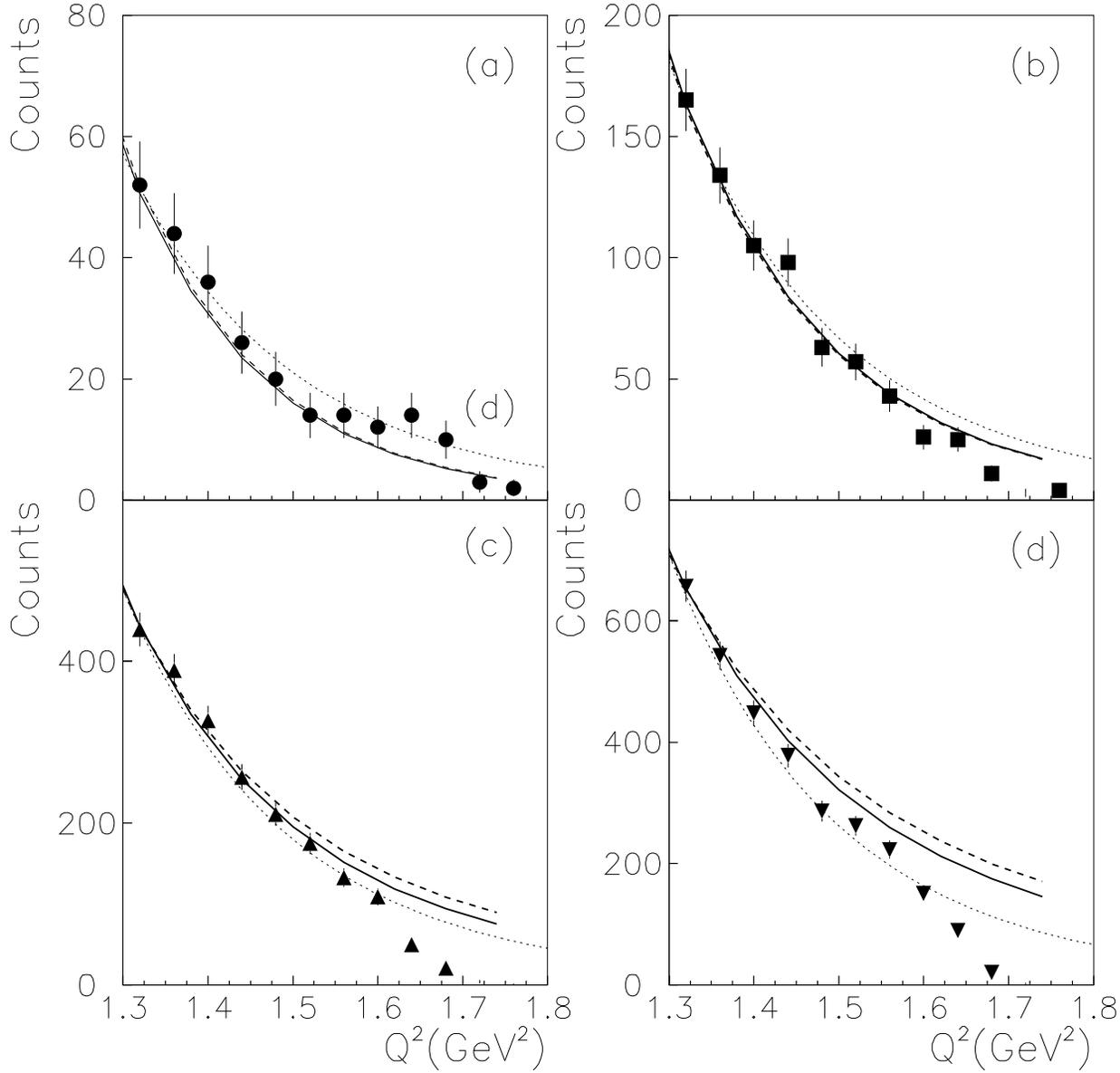}}
\end{center}
\vspace*{-3truecm}
\caption{
$Q^2$-dependence of the counting rates, corrected by the kinematical factor $\cal N$, see Eq. (\protect\ref{eq:nor}), for different bins of the
invariant mass excess $\Delta W=W-W_{th}$: 0 $\le \Delta W \le 40$ MeV (a);
40 MeV $\le \Delta W\le$ 80  MeV (b); 80 MeV $\le \Delta W \le$ 120  MeV (c); 120 MeV $ \le \Delta W\le$ 160  MeV (d). The solid line and dashed lines correspond to
different ranges of $\phi$-integration, from  \protect\cite{Re97}. The dotted line
is the pQCD prediction, with N=14 and $m^2$=1.41 GeV$^2$ see Eq. (6). All curves are normalized to data at $Q^2$=1.3 GeV$^2$.}
\label{exp}
\end{figure}
\end{document}